\documentclass[oneside,english]{amsart}
\usepackage[T1]{fontenc}
\usepackage[latin9]{inputenc}
\usepackage{geometry}
\usepackage{amsthm}
\usepackage{amstext}
\usepackage{amssymb}

\makeatletter
\numberwithin{equation}{section}
\numberwithin{figure}{section}
\theoremstyle{plain}
\newtheorem{thm}{\protect\theoremname}
  \theoremstyle{plain}
  \newtheorem{conjecture}[thm]{\protect\conjecturename}
  \theoremstyle{plain}
  \newtheorem{prop}[thm]{\protect\propositionname}
  \theoremstyle{definition}
  \newtheorem{defn}[thm]{\protect\definitionname}
  \theoremstyle{plain}
  \newtheorem{lem}[thm]{\protect\lemmaname}

\makeatother

\usepackage{babel}
  \providecommand{\conjecturename}{Conjecture}
  \providecommand{\definitionname}{Definition}
  \providecommand{\lemmaname}{Lemma}
  \providecommand{\propositionname}{Proposition}
\providecommand{\theoremname}{Theorem}

\begin{document}

\title{Proof of a conjecture by Gazeau et al. using the Gould Hopper polynomials}

\author{C. Vignat and O. Lévêque}

\address{L.T.H.I., E.P.F.L., West Lausanne, Switzerland}

\email{christophe.vignat@epfl.ch, olivier.leveque@epfl.ch}
\begin{abstract}
We prove the ``strong conjecture'' expressed in \cite{Gazeau} about
the coefficients of the Taylor expansion of the exponential of a polynomial.
This implies the ``weak conjecture'' as a special case. The proof
relies mainly about properties of the Gould-Hopper polynomials.
\end{abstract}
\maketitle

\section{Introduction}

In \cite{Gazeau}, the authors state the following conjecture: 
\begin{conjecture}
\label{Conjecture}If $\left\{ a_{i},\,\,2\le i\le p\right\} $ are
positive numbers, and with the notation $x_{n}!=\prod_{k=1}^{n}x_{k}$,
then the numbers $x_{i}$ such that 
\[
\exp\left(t+\sum_{i=2}^{p}\frac{a_{i}}{i}t^{i}\right)=\sum_{n\ge0}\frac{t^{n}}{x_{n}!}
\]
satisfy the recurrence relation
\[
x_{n}=\frac{n+1}{1+\sum_{i=2}^{p}a_{i}\frac{x_{n}!}{x_{n-i+1}!}}.
\]

\end{conjecture}
In the following, we prove this conjecture using Gould-Hopper polynomials
as defined in \cite{Gould Hopper} and some integral representations
of these polynomials as introduced in \cite{Vignat}.

\section{Preliminary Tools}

In \cite{Nieto}, Nieto and Truax consider the operator
\[
I_{j}=\exp\left[\left(c\frac{d}{dx}\right)^{j}\right]
\]
 where $c$ is a constant and $j$ an integer. They remark that, for
any well-behaved function $f$, $I_{1}$ acts as the translation operator
\[
I_{1}f\left(x\right)=f\left(x+c\right),
\]
which can also be viewed as the probabilistic expectation
\[
I_{1}f\left(x\right)=\mathbb{E}f\left(x+Z_{1}\right)
\]
 where $Z_{1}$ is the deterministic variable equal to $1.$ 

In the case $j=2,$ with $Z_{2}$ denoting a Gaussian random variable
with variance $2$, $I_{2}$ acts as the Gauss-Weierstrass transform
\[
I_{2}f\left(x\right)=\mathbb{E}f\left(x+cZ_{2}\right).
\]
 It was shown in \cite{Nieto} that this result can be extended to
any integer value of $j$ as follows:
\begin{prop}
\label{prop:Prop2}For any integer $j\ge1,$ there exists a complex-valued
random variable $Z_{j}$ such that the following representation

\begin{equation}
I_{j}=\mathbb{E}f\left(x+cZ_{j}\right)\label{eq:representation}
\end{equation}
holds.
\end{prop}
The properties of the complex-valued random variable $Z_{j}$ were
studied further in \cite{Vignat}. The only important property we
need to know here is that
\[
\mathbb{E}Z_{j}^{k}=\begin{cases}
0 & \text{if}\,\, k\ne0\,\,\mod j\\
\frac{\left(pj\right)!}{p!} & \text{if}\,\, k=pj,\,\, p\in\mathbb{N}
\end{cases}.
\]
and that, as a consequence, its characteristic function 
\begin{equation}
\mathbb{E}\exp\left(uZ_{j}\right)=\exp\left(u^{j}\right),\,\, u\ge0,\label{eq:generating function}
\end{equation}
since a straightforward computation gives
\begin{equation}
\mathbb{E}\exp\left(uZ_{j}\right)=\sum_{k=0}^{+\infty}\frac{u^{k}}{k!}EZ_{j}^{k}=\sum_{p=0}^{+\infty}\frac{u^{pj}}{pj!}\frac{pj!}{p!}=\exp\left(u^{j}\right).\label{eq:CFZj}
\end{equation}

\begin{defn}
The Gould-Hopper polynomials \emph{\cite[p.58]{Gould Hopper}} are
defined as\emph{
\begin{equation}
g_{n}^{m}\left(x,h\right)=\mathbb{E}\left(x+h^{\frac{1}{m}}Z_{m}\right)^{n}\label{eq:defGH}
\end{equation}
}and can be naturally generalized as\emph{
\begin{equation}
g_{n}\left(x,\mathbf{h}\right)=\mathbb{E}\left(x+\sum_{i=2}^{p}h_{i}^{\frac{1}{i}}Z_{i}\right)^{n}\label{eq:GHmoments}
\end{equation}
}for any vector $\mathbf{h}=\left[h_{2},\dots,h_{p}\right]$ such
that $\left\{ h_{i}\ge0,\,\,2\le i\le p\right\} .$\end{defn}
\begin{lem}
\label{lem:Lemma1}The Gould-Hopper polynomials satisfy the following
identity
\begin{equation}
g_{n}\left(x,\mathbf{h}\right)=\exp\left(\sum_{i=2}^{p}h_{i}\frac{d^{i}}{dx^{i}}\right)x^{n}.\label{eq:GHpolynomials}
\end{equation}
\end{lem}
\begin{proof}
We have
\[
\exp\left(\sum_{i=2}^{p}h_{i}\frac{d^{i}}{dx^{i}}\right)x^{n}=\prod_{i=2}^{p}\exp\left(h_{i}\frac{d^{i}}{dx^{i}}\right)x^{n}
\]
and the result follows by applying successively (\ref{eq:representation}),
we deduce the result.\end{proof}
\begin{lem}
\label{lem:Lemma2}The generating function of the Gould-Hopper polynomials
$g_{n}\left(x,\mathbf{h}\right)$ is
\[
\sum_{n=0}^{+\infty}\frac{t^{n}}{n!}g_{n}\left(x,\mathbf{h}\right)=\exp\left(xt+\sum_{i=2}^{p}h_{i}t^{i}\right).
\]
\end{lem}
\begin{proof}
From the definition (\ref{eq:defGH}), we deduce, with $Z_{2},\dots,Z_{p}$
as in Proposition \ref{prop:Prop2}, 
\begin{eqnarray*}
\sum_{n=0}^{+\infty}\frac{t^{n}}{n!}g_{n}\left(x,\mathbf{h}\right) & = & \mathbb{E}\exp\left(t\left(x+\sum_{i=2}^{p}h_{i}^{\frac{1}{i}}Z_{i}\right)\right)\\
 & = & \exp\left(xt\right)\prod_{i=2}^{p}\mathbb{E}\exp\left(th_{i}^{\frac{1}{i}}Z_{i}\right)
\end{eqnarray*}
and the result follows from (\ref{eq:CFZj}).
\end{proof}
From this lemma, we deduce that the factorial coefficients $x_{n}!$
satisfy
\[
\left(x_{n}!\right)^{-1}=\frac{g_{n}\left(x,\mathbf{h}\right)}{n!}.
\]
In order to obtain a recurrence formula for the numbers $x_{n},$
we need the following recurrence relation on the Gould-Hopper polynomials.
\begin{lem}
\label{lem:Lemma3}The Gould-Hopper polynomials (\ref{eq:GHpolynomials})
satisfy the difference equation
\[
g_{n+1}\left(x,\mathbf{h}\right)=xg_{n}\left(x,\mathbf{h}\right)+\sum_{k=2}^{p}kh_{k}\frac{n!}{\left(n-k+1\right)!}g_{n+1-k}\left(x,\mathbf{h}\right).
\]
\end{lem}
\begin{proof}
The moment representation (\ref{eq:GHmoments}) yields
\begin{eqnarray*}
g_{n+1}\left(x,\mathbf{h}\right) & = & \mathbb{E}\left(x+\sum_{i=2}^{p}h_{i}^{\frac{1}{i}}Z_{i}\right)^{n+1}=\mathbb{E}\left(x+\sum_{i=2}^{p}h_{i}^{\frac{1}{i}}Z_{i}\right)\left(x+\sum_{i=2}^{p}h_{i}^{\frac{1}{i}}Z_{i}\right)^{n}\\
 & = & x\mathbb{E}\left(x+\sum_{i=2}^{p}h_{i}^{\frac{1}{i}}Z_{i}\right)^{n}+\sum_{k=2}^{p}h_{k}^{\frac{1}{k}}\mathbb{E}Z_{k}\left(x+\sum_{i=2}^{p}h_{i}^{\frac{1}{i}}Z_{i}\right)^{n}.
\end{eqnarray*}
The first term is identified as $xg_{n}\left(x,\mathbf{h}\right)$
and the second term is computed using the following lemma.\end{proof}
\begin{lem}
The random variables $Z_{j}$ as defined in Proposition \ref{prop:Prop2}
satisfy the following Stein identity
\[
\mathbb{E}\left(Z_{k}f\left(x+\sum_{i=2}^{p}h_{i}^{\frac{1}{i}}Z_{i}\right)\right)=kh_{k}^{1-\frac{1}{k}}\mathbb{E}f^{\left(k-1\right)}\left(x+\sum_{i=2}^{p}h_{i}^{\frac{1}{i}}Z_{i}\right)
\]
for any smooth function $f.$\end{lem}
\begin{proof}
The partial derivative
\[
\frac{\partial}{\partial h_{k}}\mathbb{E}\left(f\left(x+\sum_{i=2}^{p}h_{i}^{\frac{1}{i}}Z_{i}\right)\right)=\mathbb{E}\left(Z_{k}f'\left(x+\sum_{i=2}^{p}h_{i}^{\frac{1}{i}}Z_{i}\right)\right)\frac{1}{k}h_{k}^{\frac{1}{k}-1}
\]
can also be computed from (\ref{eq:GHpolynomials}) as
\begin{eqnarray*}
\frac{\partial}{\partial h_{k}}\exp\left(\sum_{i=2}^{p}h_{i}\frac{d^{i}}{dx^{i}}\right)f\left(x\right) & = & \exp\left(\sum_{i=2}^{p}h_{i}\frac{d^{i}}{dx^{i}}\right)\frac{d^{k}}{dx^{k}}f\left(x\right)\\
 & = & \exp\left(\sum_{i=2}^{p}h_{i}\frac{d^{i}}{dx^{i}}\right)f^{\left(k\right)}\left(x\right)=\mathbb{E}f^{\left(k\right)}\left(x+\sum_{i=2}^{p}h_{i}^{\frac{1}{i}}Z_{i}\right)
\end{eqnarray*}
so that
\[
\mathbb{E}\left(Z_{k}f'\left(x+\sum_{i=2}^{p}h_{i}^{\frac{1}{i}}Z_{i}\right)\right)\frac{1}{k}h_{k}^{\frac{1}{k}-1}=\mathbb{E}f^{\left(k\right)}\left(x+\sum_{i=2}^{p}h_{i}^{\frac{1}{i}}Z_{i}\right)
\]
which is the result after replacing $f'$ by $f$ in both sides. Using
this result yields with $f\left(x\right)=x^{n}$ yields the proof
of Lemma \ref{lem:Lemma3}.
\end{proof}
We can now prove the Conjecture \ref{Conjecture} as follows: by Lemma
\ref{lem:Lemma3}, the quantities
\[
\left(x_{n}!\right)^{-1}=\frac{1}{n!}g_{n}\left(x,\mathbf{h}\right)
\]
satisfy the recurrence
\[
\left(n+1\right)\left(x_{n+1}!\right)^{-1}=x\left(x_{n}!\right)^{-1}+\sum_{k=2}^{p}kh_{k}\left(x_{n+1-k}!\right)^{-1}.
\]
Dividing both sides by $\left(x_{n}!\right)^{-1},$ and remarking
that
\[
x_{n+1}^{-1}=\frac{\left(x_{n+1}!\right)^{-1}}{\left(x_{n}!\right)^{-1}},
\]
 we deduce
\[
\left(n+1\right)x_{n+1}^{-1}=x+\sum_{k=2}^{p}kh_{k}\frac{x_{n}!}{x_{n+1-k}!}.
\]
Choosing $h_{k}=\frac{a_{k}}{k}$ and $x=1,$ we deduce the result.

We note that we have a proved slightly more general result than Conjecture
\ref{Conjecture}, namely the fact that the coefficients $x_{n}$
in the expression
\[
\exp\left(xt+\sum_{i=2}^{p}\frac{a_{i}}{i}t^{i}\right)=\sum_{n\ge0}\frac{t^{n}}{x_{n}!}
\]
satisfy the recurrence
\[
x_{n+1}=\frac{n+1}{x+\sum_{k=2}^{p}a_{k}\frac{x_{n}!}{x_{n+1-k}!}}.
\]

\end{document}